\documentclass[prl,preprint,showpacs,preprintnumbers,amsmath,amssymb,superscriptaddress,showkeys]{revtex4}

\usepackage[utf8]{inputenc}
\usepackage{amsmath}
\usepackage{amsfonts}
\usepackage{amssymb}
\usepackage{graphicx}
\usepackage{longtable}

\begin{document}

\preprint{???}

\author{F. H. Jornada}
\affiliation{%
Instituto de F\'{i}sica, Universidade Federal do Rio Grande do Sul, 91501-970 Porto Alegre - RS, Brazil
}

\author{V. Gava}
\affiliation{%
Universidade de Caxias do Sul, 95070-560 Caxias do Sul - RS, Brazil
}

\author{A. L. Martinotto}
\affiliation{%
Universidade de Caxias do Sul, 95070-560 Caxias do Sul - RS, Brazil
}

\author{L. A. Cassol}
\affiliation{%
Universidade de Caxias do Sul, 95070-560 Caxias do Sul - RS, Brazil
}

\author{C. A. Perottoni}
\affiliation{%
Universidade de Caxias do Sul, 95070-560 Caxias do Sul - RS, Brazil
}

\title{Modeling of amorphous carbon structures with arbitrary structural constraints}

\date{\today}

\begin{abstract}
In this paper we describe a method to generate amorphous structures with arbitrary structural constraints. This method employs the Simulated Annealing algorithm to minimize a simple yet carefully tailored Cost Function (CF). The Cost Function is composed of two parts: a simple harmonic approximation for the energy-related terms and a cost that penalizes configurations that do not have atoms in the desired coordinations. Using this approach, we generated a set of amorphous carbon structures spawning nearly all the possible combinations of $sp$, $sp^2$ and $sp^3$ hybridizations. The bulk moduli of this set of amorphous carbons structures was calculated using Brenner's potential. The bulk modulus strongly depends on the mean coordination, following a power law behavior with an exponent $\nu=1.51 \pm 0.17$. A modified Cost Function that segregates carbon with different hybridizations is also presented, and another set of structures was generated. With this new set of amorphous materials, the correlation between the bulk modulus and the mean coordination weakens. This method proposed can be easily modified to explore the effects on physical properties of the presence hydrogen, dangling bonds, and structural features such as carbon rings. 
\end{abstract}

\pacs{61.43.Bn, 61.43.-j, 62.20.de}

\keywords{amorphous carbon, simulated annealing, bulk modulus}

\maketitle

\section{Introduction}

Carbon is an impressively versatile chemical element. As it is found in three distinct hybridizations, $sp^3$, $sp^2$ and $sp$, each with a well-defined local topology, this element can form a variety of different allotropes. Diamond's bulk hardness and graphite's laminar softness, for instance, can each be tracked down to carbon $sp^3$ tetrahedron-like bonding and $sp^2$ planar configuration. Although these two materials exhibit quite distinctive physical and chemical properties, they represent only a small fraction of possible carbon solids.

Due to their many industrial applications, non-crystalline carbon materials have lately received attention. This class of material can be cheaply produced by Chemical Vapor Deposition (CVD)~\cite{Messina-2006} and deposited over surfaces as thin, hard films, which exhibit good biocompatibility and chemical inertness~\cite{Robertson-1902}. In these films, the amount of $sp^3$, $sp^2$ and $sp$ hybridized carbon, along with the presence or absence of hydrogen, directly influences the coating's stiffness~\cite{Fanchini-1996}. Considering the immense variety of amorphous carbon films that can be generated experimentally, many efforts were made to theoretically address how their properties vary according to changes in composition and local structure.

One of the first models to describe non-crystalline materials was elaborated by Zachariasen, who introduced the concept of continuous random network (CRN) to explain the atomic arrangement in $\mathrm{SiO_2}$ glasses~\cite{Zachariasen-1932}. Despite primarily addressing $\mathrm{SiO_2}$ structures, Zachariasen proposed that most oxide glasses could be considered as a random disposal of atoms in which there are neither bond defects nor long range crystallinity. Further advances in this field were made possible by employing computers to generate continuous random networks. Among the most successful approaches is the one by Wooten, Winer and Weaire (WWW)~\cite{Wooten-1985}. Their ingenious bond-switching method consists of randomly swapping bonds from an originally $100\%$ $sp^3$ crystalline structure. In the WWW approach, one starts with a periodic diamond supercell, and follows a cycle of bond interchanges and geometry relaxation until a fully tetrahedral amorphous carbon (ta-C) structure, also referred to as amorphous diamond (a-D), is obtained. This method is not only computationally fast and straightforward to implement, but also very successful in reproducing the experimental radial distribution function of amorphous solids~\cite{Wooten-1987}.

As computer performance improved, it became possible to generate a-C by employing Molecular Dynamics (MD). Using this strategy, it is possible to simulate either the quenching of a melted carbon liquid or the process of film deposition, and amorphous carbon structures containing $sp$, $sp^2$ and $sp^3$ hybridizations may be obtained~\cite{Kaukonen-1992, Wang-1993, Kelires-1994, Marks-2000, Marks-2002, Mathioudakis-2004}. Another approach, analogous to MD, for the computer generation of amorphous carbon structures is the activation-relaxation technique, proposed by Barkema and Mousseau~\cite{Barkema-1996}. This method also allows one to obtain a CRN containing carbon atoms in different coordination, and it is an efficient means to overcome energy barriers between metastable \textit{minima}.

Many previous works have pointed out a general trend of a-C structures to become denser and stiffer with the increase of the mean atomic coordination~\cite{Djordjevic-1997, Mathioudakis-2004}. These findings are supported by experimental data~\cite{Ferrari-1999} and by the percolation model of Phillips~\cite{Phillips-1979} and Thorpe~\cite{Thorpe-1983,Thorpe-1985}. The latter work may be seen as the theoretical basis that explains the strong dependence of elastic properties both on the mean coordination and on the presence of small rings, and it can also successfully explain CRNs' bulk modulus vanishing as the mean atomic coordination $\overline{z}$ approaches $2.4$.

Computer methods to generate a-C have recently included the use of increasingly sophisticated Hamiltonians (such as \textit{ab initio}) to more accurately simulate the dynamics of carbon atoms~\cite{Galli-1989, Marks-1996a, McCulloch-2000, Han-2007}. Likewise, many of these algorithms focus on simulating the experimental process that originates a particular amorphous system. This may be seen as a \textit{top-down} strategy, as for each new a-C material produced, it is necessary to perform an MD that resembles the corresponding experimental conditions. Only after that can the material's properties be estimated. For instance, if one were to answer the effect of a particular local structure on the amorphous carbon properties, various MDs might possibly have to be performed until a certain simulation yielded the desired microscopic structure.

Later, with the discovery of new forms of a-C, the influence of some local scale features on the properties of amorphous systems was pointed out~\cite{Lyapin-2000, Hong-2009}. The intrinsic complexity of dealing with these features, such as the presence of rings, the proportion of $sp$, $sp^2$ and $sp^3$ hybridized atoms, and the size of clusters, may explain the difficulty to obtain an universal relationship among density, bulk modulus and mean coordination~\cite{Lyapin-2000}. In order to efficiently study these aspects, it is important to develop a method to generate amorphous carbon structures, as neither WWW's algorithm nor MD can easily generate a-C with generic structural constraints. The former because it is limited to $sp^3$ hybridized carbon, and the latter because it follows a \textit{top-down} approach and offers no direct way to control the presence of those features.

In order to tackle these difficulties, we introduce a scheme for the computer generation of a-C structures which follows a \textit{bottom-up} strategy. Relying on the algorithm of Simulated Annealing (SA)~\cite{SA}, we propose a Cost Function (CF) that aims not at simulating realistic atomic dynamics -- as done in~\cite{Blaudeck-2003} -- but rather at exploring several metastable configurations that meet some desired criteria. Our aim is to introduce something of a ``theoretical workbench'' to flexibly simulate a-C. One of our main interests is to employ this technique to calculate the bulk modulus' dependence on the fraction of $sp$, $sp^2$ and $sp^3$ hybridized carbon, and to perhaps find an amorphous structure more incompressible than diamond.

The paper is organized as follows: First, in the computational strategy section, we detail the proposed Cost Function. Next, we present the results of simulations with 512 carbon atoms, using periodic boundary conditions, in which we calculate the bulk modulus of several amorphous structures spawning several possible combinations of $sp^3$, $sp^2$ and $sp$ carbon and also compare the calculated radial distribution functions with the literature. We then show that we can modify our CF to force the creation of clusters of atoms with the same hybridization, and thus explore their effect on the bulk modulus. We conclude by discussing the advantages of the approach described in this paper, as well as possible extensions that can be proposed to deal with other problems.

\section{Computational Strategy}

Following the idea of the SA technique, we have developed a fast and customizable CF which guides the exploration of different atomic configurations. Our Cost Function is composed of two parts, the first a computationally simple potential, and the second depending on the desired constraints for the CRN. The CF is required neither to be a continuous function nor to yield a realistic value for structures far from a metastable situation; its only requirement is to have an arbitrary low value for geometrically stable configurations. Thus, the problem of finding a certain amorphous material that meets some arbitrary constraints can be transformed to the problem of finding a reasonably low minimum of this Cost Function using SA. 

Considering the potential part of the CF, a simple harmonic-like approximation was employed. As we wished our model to be as computationally fast as possible, carbon atoms were considered to be either bonded or non-bonded, with a cut-off distance $r_{c}$ of $2.0$ \AA{}. This value is somewhat arbitrary, as long as it is smaller than the typical second-neighbor distance, but we found that shrinking it too much allows non-bonded atoms to remain too close. Thus, the potential term of the Cost Function is written as

\begin{align}
\phi_V = v_r \sum_{ r_{ij} } (r_{ij} - r^\ast_{c(i) c(j)} ) ^2 
+ v_a \sum_{ \theta_{ijk} } (\theta_{ijk} - \theta^\ast_{c(j)} )^2 \notag \\
+ v_t \sum_{\mathbf{u}_i,\mathbf{u}_j} [ 1- (\mathbf{u}_i \cdot \mathbf{u}_j)^2 ]
\label {eq:phi_V}
\end{align}

The first sum is over all bonds $r_{ij}$ and expresses the stretching energy relative to the equilibrium distance $r^\ast_{c(i) c(j)}$  between atoms $i$ and $j$ (having coordination $c(i)$ and $c(j)$ respectively). The second sum comprises all $\theta_{ijk}$ angles having a common $j$ center, where $\theta^\ast_{c(j)}$ denotes the equilibrium angle, which in our approximation depends only on the hybridization $c(j)$ of atom $j$. The last sum considers only connected $sp^2$ centers, and it constitutes the torsional energy of having two $sp^ 2$ planes, with normal vectors $\mathbf{u}_i$ and $\mathbf{u}_j$, nonparallel. In all summations, repeated terms are not counted. Also, the constants $v_r$, $v_a$ and $v_t$ merely set the relative strength of the radial, angular and torsion terms for the total $\phi_V$.

The equilibrium quantities do not need to be found with great precision. Since our objective is simply to obtain a structure obeying a given set of constraints, there can be small geometrical distortions, all of which can be removed in a further relaxation by using a more realistic Hamiltonian. Thus, the following approximations were made: if two bonded atoms are both $sp^3$, their equilibrium distance $r^\ast_{4 4}$ is the same found in diamond; for $sp^2$-$sp^2$ bonds, equilibrium interatomic distance $r^\ast_{3 3}$ is that found in graphite and, in the case of $sp$-$sp$ bonds, one takes the equilibrium distance of the triple bond in 2-butyne for $r^\ast_{2 2}$~\cite{xray}. For a pair of bonded carbon atoms with different hybridizations, say, $c'$ and $c''$, the equilibrium distance $r^\ast_{c' c''}$ is simply is the average $(r^\ast_{c' c'}+r^\ast_{c'' c''})/2$. Finally, if an atom has a coordination $c'>4$, the same value as for four-fold atoms is assumed ($r^\ast_{c' c'}=r^\ast_{4 4}$). Likewise, the distance $r^\ast_{1 1}$ between singly-bonded atoms is the same as for $sp$-$sp$.

The way we presented $\phi_V$ alone will not yield any bondings, as linked atoms will always increase the system's energy. In order to correct this and to control the material hybridizations, we introduce another term in the Cost Function. This term, here referred to as Coordination Cost, allows one to set how many atoms should be $sp^3$, $sp^2$ and $sp$:

\begin{equation}
\phi_C = \sum_{c'} \epsilon_{c'} | n_{c'} - n^\ast_{c'} |
\label {eq:phi_C}
\end{equation}

The sum is over all possible $c'$ coordinations, where $n_{c'}$ is the number of $c'$-coordinated atoms, and $n^\ast_{c'}$ is a parameter that sets how many atoms should be $c'$-coordinated. This way, each constant $\epsilon_{c'}$ sets a cost for a configuration having a wrong number $|n_{c'} - n^\ast_{c'}|$ of $c'$-coordinated centers. Clearly, we must have $\sum_{c'} n^\ast_{c'} = \sum_{c'} n_{c'} = N$, where $N$ is the total number of atoms. Taking the absolute value of $n_{c'} - n^\ast_{c'}$ rather than squaring it has shown some advantages, such as the Cost Function exhibiting a sharper minimum. Moreover, it naturally makes the CF an extensive function, so that constants do not have to be altered as the number of atoms in the simulation box changes.

Finally, the Cost Function $\Phi$ is simply defined as a linear combination of the previous terms:

\begin{equation}
\Phi = \lambda_{V} \phi_V + \lambda_{C} \phi_C,
\label {eq:Phi1}
\end{equation}

\noindent where $\lambda_{V}$ and $\lambda_{C}$ are constants. With the former definitions, bonded atoms are stable, provided that their linking decreases the number of atoms with wrong coordinations. By putting $N$ atoms in a cubic cell with periodic boundaries and setting how many should be $sp^3$, $sp^2$ and $sp$ hybridized (\textit{i.e.}, fixing $n^\ast_4$, $n^\ast_3$ and $n^\ast_2$), a CRN can be obtained as the set of atomic positions which minimizes $\Phi$. We expect our CF to possess many metastable \textit{minima}, and we employed the Simulated Annealing (SA) algorithm to optimize $\Phi$.

Despite being SA a technique aimed at finding the global minimum of complex hypersurfaces, we are not necessarily interested in using this technique to its full extent. Consider, for instance, the generation of fully $sp^3$ carbon CRNs. The global minimum of the CF in this case is crystalline diamond, which is clearly not the solution we are looking for. Accordingly, we propose the use of the SA algorithm to find deep minima of the Cost Function that comply with a certain set of constraints. Whether or not this minimum is the global minimum of the Cost Function is not our concern.

We devised the optimization algorithm the following way: Each step in the SA scheme constitutes randomly displacing one atom inside a cube of side length $0.4$ \AA{}, changing the CF by $\Delta \Phi$. Even though system-wide movements could be implemented, individual movements have shown a great advantage, as $\Phi$ only changes locally in this case. So, the recalculation of $\phi_V$, which is the most numerically intensive task in our CF, only has to be evaluated in a small radius around the displaced atom. This way, the computational complexity of $\Delta \Phi$ approximately does not scale with the system size. As in classical SA, each movement is accepted stochastically according to the weighing factor $e^{-\beta \, {\Delta \Phi}}$, where $\beta$ is the inverse of the fictitious temperature $T$. In addition to atomic displacements, the system periodically undergoes a random expansion or contraction, so that one does not need to set the final density \textit{a priori}. Although $\Delta \Phi$ can not be evaluated locally in this case, the number of atomic displacements between full system scalings are such that these resizings do not compromise the algorithm speed.

A key component to successfully finding a minimum for $\Phi$ is the determination of the optimal annealing scheme~\cite{Johnson-1989}. Following Christoph and Hoffmann~\cite{Christoph-1993}, we decreased $T$ using a power law. In order to further control the annealing, we separated it in three regions, and following Johnson \textit{et al.}~\cite{Johnson-1989}, we fixed the initial and final acceptance rates instead of the temperatures, as the former proved to be less sensitive to changes in the CF. The initial and final acceptance rates and number of steps assigned to each annealing region were found by minimizing the width of the angular distribution function for a-D, and results are summarized in Table \ref{tab:regions}.

\begin{table}
\caption{\label{tab:regions} Parameters used in the three-regions annealing scheme: initial ($P_i$) and final ($P_f$) acceptance rate and the fraction of the total steps assigned to each region.}
\begin{ruledtabular}
\begin{tabular}{c c c c}
Region & Steps & $P_i$ & $P_f$ \\ \hline 
$1$ & $10\%$ & $95\%$ & $70\%$ \\
$2$ & $80\%$ & $70\%$ & $30\%$ \\
$3$ & $10\%$ & $30\%$ & $5\%$ \\ 
\end{tabular}
\end{ruledtabular}
\end{table}

The constants in Eqs.~\eqref{eq:phi_V}-\eqref{eq:Phi2} were determined as follows. Medium-sized (256 atoms) mixed-coordination amorphous carbon networks were generated, and the resulting structures were further relaxed using Molecular Dynamics  and Brenner's interatomic potential~\cite{Brenner-2002} using GULP~\cite{Gulp}. Parameters were thus varied to minimize the final structures' energies and coordination errors (\textit{i.e.}, the numbers $n_{c'}$ of $c'$-coordinated atoms should be as close as possible to the defined number $n_{c'}*$). The optimal constants are shown in Table~\ref{tab:constants}. Also, we found $\lambda_V / \lambda_C \approx 0.75$ to be appropriate for most simulations. One should note that these constants could not be determined with great precision due to the large statistical fluctuations in the methodology. Fortunately, precise values are not sought, since small stresses in the CRNs could be eliminated \textit{a posteriori} by a supplemental Molecular Dynamics using more sophisticated interatomic potentials, such as Brenner's potential~\cite{Brenner-2002}, or even \textit{ab initio} calculations. 

\begin{table*}
\caption{\label{tab:constants} Parameters for Eqs.~\eqref{eq:phi_V}-\eqref{eq:Phi2}.}
\begin{ruledtabular}
\begin{tabular}{l c c c c c c c c c c c c c}
Constant & $\lambda_V$ & $\lambda_C$ & $v_r$ & $v_a$ & $v_t$ \\
Value    & $1.0$       & $2.5$  & $5.0$ & $3.0$ & $1.5$ \\ \hline
Constant & $\epsilon_0$ & $\epsilon_1$ & $\epsilon_2$ & $\epsilon_3$ & $\epsilon_4$ & $\epsilon_5$ & $\epsilon_{j (j\ge 5)}$ \\
Value    & $10.0$       & $5.0$        &  $2.0$       & $1.5$        & $1.0$        & $10.0$       & $10^j$  \\ \hline
Constant & $r_{11}^\ast$  & $r_{22}^\ast$  & $r_{33}^\ast$   & $r_{44}^\ast$   & $r_{jj (j\ge 4)}^\ast$ \\
Value $(\AA)$    & $1.2$ & $1.2$ & $1.42$ & $1.54$ & $1.54$ \\ \hline
Constant & $\theta_{2}^\ast$  & $\theta_{3}^\ast$  & $\theta_{4}^\ast$    & $\theta_{j (j\ge 4)}^\ast$ \\
Value    & $180^{\circ}$ & $120^{\circ}$ & $109.4^{\circ}$ & $109.4^{\circ}$ 
\end{tabular} 
\end{ruledtabular}
\end{table*}

As a next step, in order to validate our CF, we first generated a 64-atom $100\%$ $sp^3$ CRN. We used this CRN and a diamond structure as references to build 118 other CRNs, each of them constructed as a linear combination of the two reference structures. More specifically, denoting $\mathbf{r}_i^d$ and $\mathbf{r}_i^a$ the positions of the $i$th atom of the diamond and the amorphous structure respectively, each interpolated CRN was defined according to $\mathbf{r}_i(u)  = (1-u)\;\mathbf{r}_i^d + u\;\mathbf{r}_i^a$, where $u$ is an interpolation parameter.

\begin{figure}
\centering
\includegraphics[scale=1]{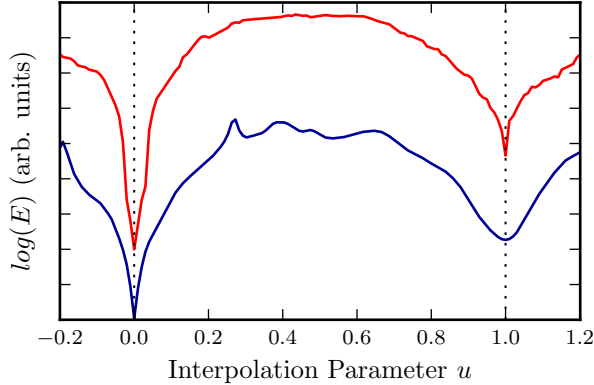}
\caption{\label{fig:pots} (Color online). Comparison of the energy calculated using our Cost Function (top red curve) and Brenner's~\cite{Brenner-2002} potential (bottom blue) for a set of structures. Top curve has been displaced vertically for clarity. The left minimum at $u=0$ corresponds to a diamond structure, and the right one at $u=1$ to a $sp^3$ amorphous material.}
\end{figure}

For each CRN, we calculated the energy using our Cost Function and Brenner's potential, and plotted the results in Fig.~\ref{fig:pots}. It is clear that both models should yield high values for unstable materials (\textit{i.e.}, for $u$ far from $0$ or $1$), and that our potential should only partially reproduce the true energy surface. Nevertheless, our simplified Cost Function very much resembles the computationally more expensive Brenner's potential and, particularly, it agrees with it on the position of the two \textit{minima}, which is the key feature for the Simulated Annealing scheme to identify a metastable CRN.

\section{Results and Discussion}
The procedure we are proposing for the generation of customized continuous random networks was first applied to map the bulk modulus' dependence on the fraction of $sp$, $sp^2$ and $sp^3$ hybridized carbon. To do so, we generated $45$ amorphous structures, each containing $512$ atoms and having a different proportion of the possible atomic coordinations. These proportions were chosen so as to homogeneously cover a ternary graph mapping all possible coordinations. It took $2\times 10^8$ iterations ($4$ hours\footnote{Simulations were carried out in a 32-bits, single core Intel\copyright Celeron\copyright CPU with 2.66GHz clock and 1Gb RAM.}) for each SA simulation. After these structures were generated, they were submitted to Molecular Dynamics simulations using Brenner's potential to minimize non-physical features, such as distorted angles and distances. The MD also ensures that each structure stays in a relatively low energy metastable configuration.

Each MD simulation was carried out at a low temperature of $50$ K in order to preserve the main features of the CRN generated by SA. The structures that were modified the most by the Molecular Dynamics process had roughly $50\%$ $sp$/$sp^3$ carbon atoms, but little or no $sp^2$ centers. In the worst case, a structure with $\overline{z}=2.87$ had $14.65\%$ of its $sp^3$ and $2.15\%$ of its $sp$ atoms turned into the $sp^2$ form. On the other hand, the most $sp^3$-rich structure, having final mean coordination $\overline{z}=3.98$, had less than $1.5\%$ change in its hybridization due to the MD relaxation.

\begin{figure}
\centering
\includegraphics[width=200pt]{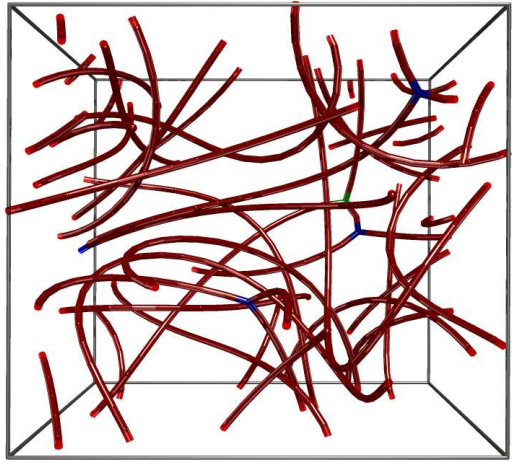}\\(a)\\
\includegraphics[width=200pt]{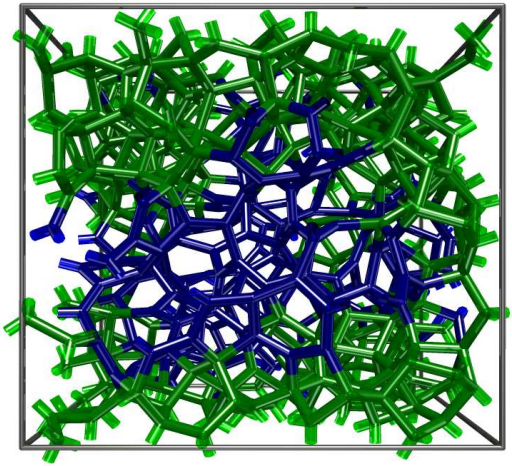}\\(b)\\ 
\includegraphics[width=200pt]{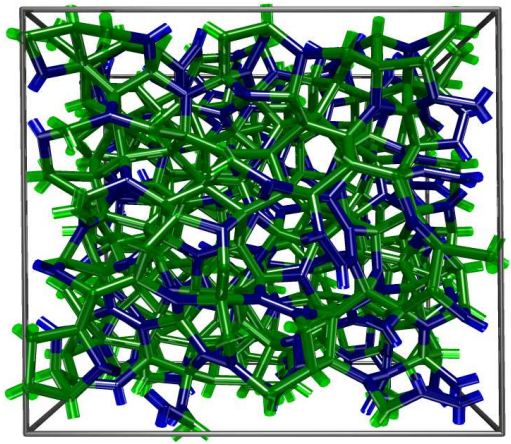}\\(c)
\caption{\label{fig:structures} (Color online). Example of some generated structures. The following color scheme was used: $sp^3$ atoms are shown in green, $sp^2$ ones in blue and $sp$ in red. (a) A $sp$ rich amorphous carbon network. (b) A mixed $sp^2$/$sp^3$ structure with $\lambda_H=1.5$. Note that the CRN segregates $sp^2$ and $sp^3$ atoms in two phases due to the heterogeneity cost. (c) Another mixed $sp^2$/$sp^3$ structure but with $\lambda_H=0$. There are no visibly distinct phases in this case.}
\end{figure}

Due to the high computational cost required to perform a full MD, only the equilibration phase was performed. Each MD isobaric simulation was performed for $5$ ps using a $0.1$ fs time step. Afterwards, each structure was submitted to a full Hessian-driven geometry relaxation, so that the elastic moduli could be calculated. We could have also estimated the the elastic moduli directly using the volume fluctuations of the MD, but it would have required a much longer time period. Both the MD simulations and the Hessian-driven geometry relaxations were performed using GULP~\cite{Gulp}. Fig.~\ref{fig:structures} shows some CRNs generated, including a $sp$ rich carbon network, which may be quite difficult to obtain using Molecular Dynamics under conventional approaches. One should note, however, that Brenner's potential does not includes van der Waals forces, which should be quite important in determining the geometries and elastic properties of these low-density $sp$ carbon structures.

As an example of the flexibility of our approach to generate a-C, we also proposed another term to the Cost Function to penalize binded atoms with different coordinations. This term was proposed to segregate carbon atoms having different hybridizations for main reasons. The first is purely theoretical, as, at least in principle, $sp^2$ centers embedded in a rigid $sp^3$ matrix should not influence the bulk modulus much, whereas if these threefold atoms form a segregate phase they might make the material considerably more compressible. This $sp^2$ segregation is also motivated by the existence of experimental carbon materials containing $sp^3$-rich phases and graphite-like $sp^2$-rich regions~\cite{Lau-2008}. The second reason is that it has been pointed out that the microstructure of hydrogenated a-C, mainly the size and shape of $sp^2$ clusters, plays an important role in determining the electronic properties of these materials~\cite{Theye-2002}. So, we proposed an heterogeneity term $\phi_H$,

\begin{equation}
\phi_H = \sum_{r_{ij}} (1-\delta_{c(i), c(j)})
\end{equation}

\noindent where we are again summing over all possible bonds, and $c(i)$ is the coordination of the center $i$, and $\delta$ is the Kronecker delta function.

With the addition of this term, the cost function becomes:
\begin{equation}
\Phi = \lambda_V \phi_V + \lambda_C \phi_C + \lambda_H \phi_H
\label{eq:Phi2}
\end{equation}

In our simulations, we used $\lambda_H$ = $1.5$, which was the smaller number for which atoms with different coordinations would be visibly segregated. This by no means precluded that a small number of atoms with other coordinations were found in the segregated phases. One may control how pure in terms of hybridization these regions are by varying $\lambda_H$. Using this new $\Phi$, we generated another set of $45$ structures, later employing Brenner's potential as well. In order to distinguish from the other set of structures, we shall call the latter a-Cs (generated with $\lambda_H>0$) \textit{heterogeneous structures}, and the former ones (with $\lambda_H=0$) \textit{homogeneous structures}. Fig.~\ref{fig:structures} shows the effect of including $\phi_H$ into the Cost Function by showing two CRNs generated with different values of $\lambda_H$ but having similar amounts of $sp^2$ and $sp^3$ centers. 

The possibility of creating homogeneous and heterogeneous structures exemplifies the flexibility of our method to generate amorphous structures. By adding a simple and intuitive term to $\Phi$, it is possible to generate CRNs with quite different characteristics. This same approach can be employed to generate CRNs having other microscopic features. For instance, a term may be added to increase the energy of a CRN if $n$-fold carbon rings are present. Also, the constant $n_1^\ast$ may have a non-zero value, and centers with only one bond may be readily mapped into hydrogen atoms or dangling bonds.

\begin{figure}
\centering
\includegraphics[width=250pt,bb=0 0 360 312,keepaspectratio=true]{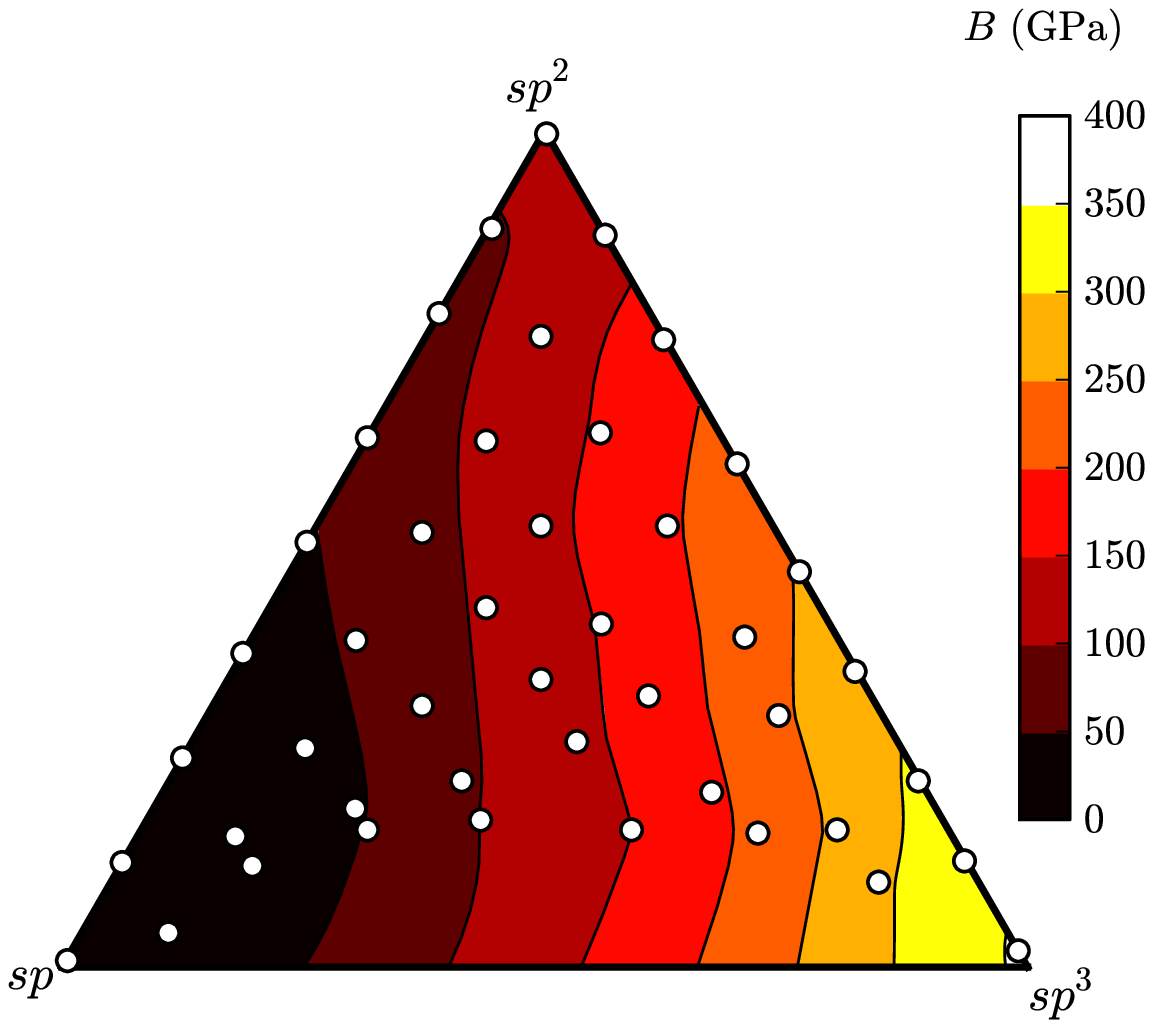}
\includegraphics[width=250pt,bb=0 0 360 312,keepaspectratio=true]{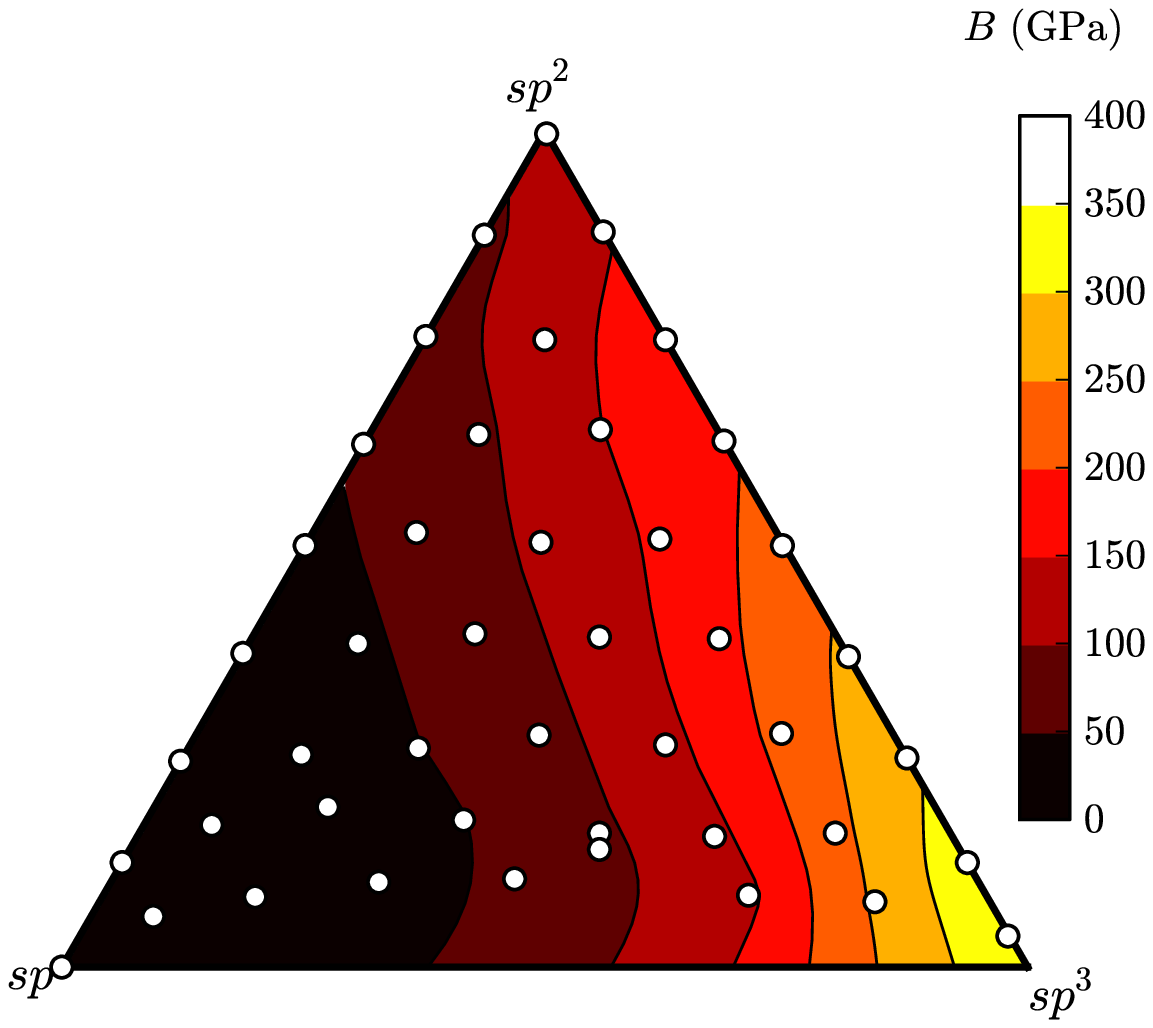}
\caption{\label{fig:ternary} (Color online).
Bulk modulus dependency on carbon hybridization. In each triangle, the lower left vertex represents a $100\%$ $sp$ structure (having a mean coordination $\overline{z}=2$), the top vertex a $100\%$ $sp^2$ structure (with $\overline{z}=3$), and the lower right vertex a $100\%$ $sp^3$ material (with $\overline{z}=4$). Points lying on the same vertical line have the same mean coordination. Top: No constraint was imposed on the heterogeneity ($\lambda_H=0$). The bulk modulus varies little along any vertical line, suggesting that it may be well described by the mean coordination. Bottom: Heterogeneous structures generated with $\lambda_H=1.5$. The mean coordination does not dictate the bulk modulus as well as in the previous case, since $B$ varies along vertical lines.}
\end{figure}

\begin{figure}
\centering
\includegraphics[scale=1]{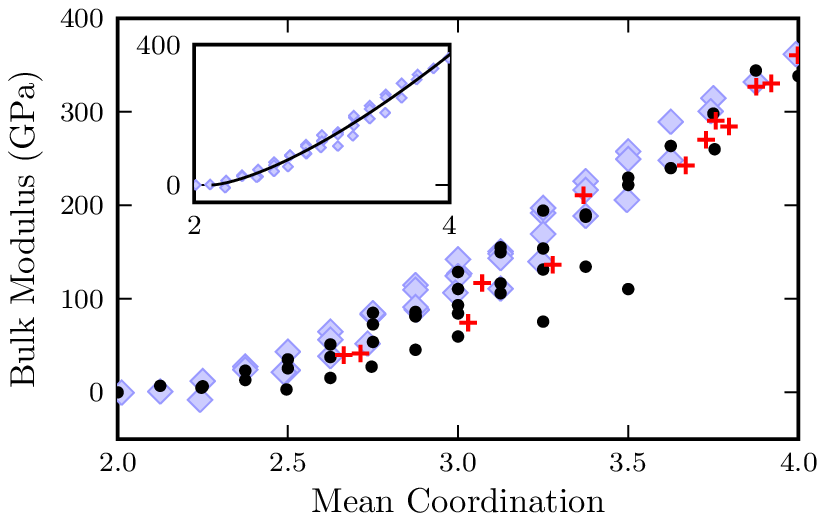}
\caption{\label{fig:bulk_2d} (Color online). Variation of the bulk modulus as a function of the mean coordination. Light diamonds (dark circles) represent data from CRNs generated by SA without (with) heterogeneity cost. For comparison, crosses show tight-binding results from~\cite{Mathioudakis-2004}. Inset: solid line representing the power law fit to data for homogeneous CRNs.}
\end{figure}

One important question that can be readily answered using our algorithm is how the bulk modulus varies with the atomic coordinations. Although it has been pointed out that the bulk modulus should depend mainly on the mean coordination, no previous method exists to generate a-C with predetermined fractions of $sp^3$, $sp^2$ and $sp$ carbon, so that this aspect has not yet been fully studied. We show in Fig.~\ref{fig:ternary} the resulting bulk modulus as a function of the fraction of $sp^3$, $sp^2$ and $sp$ carbon, for both the cases of homogeneous and heterogeneous structures\footnote{The graphics were generated using a custom-made module for pylab, and may be obtained free of charge from fjornada@if.ufrgs.br.}. We also plot the bulk modulus versus the mean coordination for the homogeneous case in Fig.~\ref{fig:bulk_2d}. The largest bulk modulus we found was  $362$~GPa, which is lower than that calculated for crystalline diamond using Brenner's potential ($442$~GPa, the same as the diamond's experimental bulk modulus~\cite{kittel}).

For the homogeneous case, the bulk modulus depended mainly on the mean atomic coordination, with a Spearman's rank correlation coefficient~\cite{Spearman-1904} $\rho=0.98$ -- supporting previous studies also pointing out this trend~\cite{Phillips-1979, Thorpe-1983, Thorpe-1985, Mathioudakis-2004}. For heterogeneous networks, the dependence on the mean correlation diminished a little, with $\rho=0.9$. However, considering only the region with mean coordination $2.5<\overline{z}<3.5$, both correlations drop to $\rho=0.94$ and $\rho=0.83$, respectively. We explain the larger decrease of $\rho$ for heterogeneous structures structures this way: since $sp$ hybridized atoms form floppy~\cite{Thorpe-1983} phases with null bulk modulus, some heterogeneous CRNs, such as one made of $50\%$ $sp^3$ and $50\%$ $sp$ carbon, will have a very small bulk modulus due to the large floppy region. Such small bulk modulus will not be observed in a $100\%$ $sp^2$ network, even though both structures have the same mean coordination. If we put $\lambda_H=0$ and let homogeneous structures form, the $sp$ carbon will not segregate, but it will be incorporated between $sp^3$ centers. Thus, there will be no large floppy regions.

The bulk modulus is also plotted as a function of the mean coordination (Fig.~\ref{fig:bulk_2d}). Following previous studies~\cite{Thorpe-1985, Djordjevic-1997, Mathioudakis-2004}, we fitted a power law to the bulk modulus data for the set of homogeneous CRNs,

\begin{equation}
B(\overline{z}) = B_0 \left( \overline{z} - \overline{z}_f \right)^{\nu}
\label{eq:bulk}
\end{equation}

We found the phase transition from rigid to floppy states~\cite{Thorpe-1983} to occur at mean coordination $\overline{z}_f = 2.10 \pm 0.11$, with $B_0 = 140 \pm 26$ GPa and $\nu = 1.51 \pm 0.17$. These results, particularly the exponent, are close to those reported by~\cite{Thorpe-1985, Djordjevic-1997, Mathioudakis-2004}, as compared in Table \ref{tab:bulk}. The slight deviation for $\overline{z}_f$ can be explained by the size of the simulation cell: Even for relatively large cells containing $512$ atoms, there is a chance that a non-floppy carbon chain of $sp^2$ or $sp^3$ atoms will percolate the periodic cell. This was not observed by Mathioudkis \textit{et al.} -- whose results were extrapolated for mean coordination bellow $\overline{z}=2.68$ -- nor by He and Thorpe~\cite{Thorpe-1985} and Djordjevic and Thorpe~\cite{Djordjevic-1997}, because of a limitation of the bond depleting method which causes the simulation cell to collapse for small $\overline{z}$.

\begin{table}
\caption{\label{tab:bulk} Comparison of the fitted parameters for Eq.~\eqref{eq:bulk}.}
\begin{ruledtabular}
\begin{tabular}{l c c}
Reference & $\overline{z}_f$ & $\nu$ \\ \hline
He and Thorpe~\cite{Thorpe-1985} & $2.4$  & $1.5 \pm 0.2$ \\
Djordjevic and Thorpe~\cite{Djordjevic-1997} & $2.4$  & $1.4$ \\
Mathioudakis \textit{et al.}~\cite{Mathioudakis-2004} & $2.33$  & $1.5 \pm 0.1$ \\
This Work & $2.10 \pm 0.11$  & $1.51 \pm 0.17$
\end{tabular} 
\end{ruledtabular}
\end{table}

It must be pointed out that long-range effects may be taken into account after the amorphous structures are generated. In our case, we employed the Brenner potential, which does not include such interactions, for the relaxation and calculation of the bulk moduli. It is reasonable to assume that the long $sp$ chains are weakly binded by dispersive forces, so that the bulk modulus of floppy networks do not vanish completely. So, it is quite possible that using a potential model for the calculation of the elastic properties that includes van der Waals interactions would yield higher bulk moduli for low $\overline{z}$.

\begin{figure}
\centering
\includegraphics[scale=1]{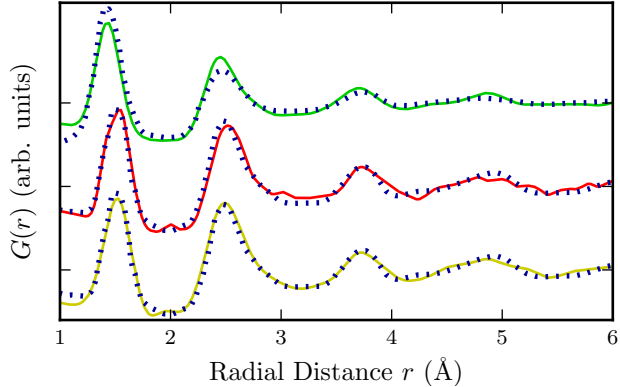}
\caption{\label{fig:rdf}(Color online). Reduced Radial Distribution Function $G(r)$. Top curves: sputtered a-C~\cite{Li-1990} (dotted line) and best fit ($100\%$ $sp^2$ structure, green solid line). Middle curves: ta-C~\cite{Gilkes-1995} (dotted line) and best fit (heterogeneous $50\%$ $sp^2$ and $50\%$ $sp^3$ structure, red solid line). Bottom curves: the same experimental curve was used~\cite{Gilkes-1995}, but the theoretical curve could also scale in the $r$ axis. The yellow curve represents a CRN containing $90\%$ $sp^3$, $10\%$ $sp^2$ and $10\%$ $sp$ carbon atoms.}
\end{figure}

After generating the CRNs, the set of $90$ structures may be seen as the initial set of an expanding database that may be used, for instance, to extract structural information from experimental results. As an example of this application, we compared the calculated radial distribution functions (RDF) for the set of CRNs with results from the literature in Fig.~\ref{fig:rdf}. Using all our available structures, we searched for CRNs that would best reproduce the RDF of some experimental materials: one sputtered a-C~\cite{Li-1990} and one ta-C~\cite{Gilkes-1995}. The first experimental a-C was prepared by rf sputtering, while the ta-C was grown using filtered cathodic arc. Using least-squares fitting, we found that a $88\%$ $sp^2$ and $12\%$ $sp$ homogeneous structure best reproduced the RDF of the sputtered a-C, while a heterogeneous $50\%$ $sp^3$/$sp^2$ structure best described the ta-C one. Furthermore, by adding an additional degree of freedom to scale the $r$ variable, the structure that best described the ta-C one was CRN containing $80\%$ $sp^3$, $10\%$ $sp^2$ and $10\%$ $sp$ carbon atoms.

The fit is not optimal: The mean coordinations of the experimental structures were reported to be $3.34$ and $3.9$, while the fitted mean coordination were $2.9$ and $3.5$ ($3.74$ if we add the additional degree of freedom). This error could be related to the potential used in the relaxation process, or to the number of iterations used to generate the structures, which in turn control their angular and bond distribution widths. However, even though the CRNs were not generated for this purpose, the comparison between experimental and theoretical RDFs was performed to show that the generated CRNs indeed present similarities with the experimental structures, to such an extent that they are able to reproduce experimental RDFs. It should be noted that it has never been in the scope of this work to present a method to extract structural information or create structures based on given experimental RDFs. There are specialized methods for this tasks, such as Reverse Monte Carlo~\cite{Mcgreevy-1988} and Hybrid Reverse Monte Carlo~\cite{Opletal-2002}, which are much more efficient to extract information from experimental RDFs, but not to generate CRNs having particular coordination or structural constraints.

Finally, we generated a set of structures to test the performance of the algorithm to generate $100\%$ $sp^3$ structures.  CRNs having 64, 128, 256 and 512 atoms were generated, and two structures were created for a given number of atoms. After relaxation using the same strategy as before (molecular dynamics followed by Hessian-driven relaxation), the calculated angular widths ranged from $11.8^{\circ}$ to $13.9^{\circ}$, with a mean value of $13.2^{\circ}$. We did not observe a significant correlation between the angular width and the number of atoms, and the angular distribution was relatively symmetric with an average of $109.07^{\circ}$. For comparison, high-quality tetrahedral networks having an angular width of $9.19^{\circ}$ have been assembled using a modified version of the WWW algorithm by Barkema and Mousseau~\cite{Barkema-2000}. Also, a reduction of the angular distribution width may be possible by increasing the number of steps during the SA or by extending the relaxation process after the structure is generated. So, although our strategy is not optimized for a-D as other methods, it is flexible enough to generate both a-D and a-C with various hybridizations and structural constraints.

\section{Conclusion}

In this paper, we described the computational creation of carbon CRNs employing the Simulated Annealing algorithm. We proposed a numerically simple Cost Function able to yield extremely different amorphous materials. As an example of the capabilities of our algorithm, we generated amorphous structures spawning nearly all possible combinations of $sp^3$, $sp^2$ and $sp$ hybridized centers, and then calculated their bulk moduli using Brenner's potential~\cite{Brenner-2002}. We were also able to easily modify our CF to create heterogeneous materials, in which atoms with the same hybridization tend to segregate. With the set of homogeneous structures, we observed a phase transition from floppy to rigid networks, and a power-law fitting of the bulk modulus dependency on mean atomic coordination was in close agreement with the literature. However, we noticed that the mean coordination $\overline{z}$ did not correlate with the bulk modulus of heterogeneous networks as well as it did for homogeneous ones. This indicates that the heterogeneity may play a very important role in dictating the elastic properties of a-C.

The strategy we described is completely universal and customizable, and modifications can be easily made to include other chemical elements, such as hydrogen, and to control the presence of other features, such as rings and dangling bonds. Once CRNs with particular features are generated, their physical properties can be estimated using more sophisticated Hamiltonians, including \textit{ab initio} calculations, whenever it is computationally feasible. Further extension of this approach to include microstructural constraints in the process of generating CRNs (such as those possibly responsible for ultrahigh hardness in~\cite{Dubrovinskaia-2005}) depends only on the availability of suitable computational resources.

\section{Acknowledgements}

The authors thank Ricardo Vargas Dorneles and Gunther Johannes Lewczuk Gerhardt, who granted access to the computer clusters where most of the calculations presented in this paper were performed, and John Muller and Veronica Gouvea for reviewing this manuscript. This work was partially supported by the Brazilian agency CNPq (Conselho Nacional de Desenvolvimento Cient\'{\i}fico e Tecnol\'{o}gico). We would also like to thank Universidade de Caixas do Sul for the computer resources (GridUCS).

\bibliography{crns}

\end{document}